\documentclass[11pt,a4paper]{article}
\usepackage[margin=27mm]{geometry}
\usepackage[T1]{fontenc}
\usepackage[utf8]{inputenc}
\usepackage{lmodern}
\usepackage{amsthm}
\usepackage{enumitem}
\usepackage{microtype}
\usepackage{fancyhdr}
\newtheorem{theorem}{Theorem}[section]
\newtheorem{proposition}[theorem]{Proposition}
\newtheorem{lemma}[theorem]{Lemma}
\newtheorem{corollary}[theorem]{Corollary}
\theoremstyle{definition}
\newtheorem{definition}[theorem]{Definition}
\newtheorem{example}[theorem]{Example}
\theoremstyle{remark}
\newtheorem{remark}[theorem]{Remark}

\usepackage{amsmath,amssymb}
\usepackage{cite}
\usepackage{url}
\usepackage{hyperref}
\hypersetup{hidelinks}

\begin{document}

\title{Positivity Improving Quantum Markov Operators}
\vskip1cm
\author{Luigi Accardi\\Centro Interdipartimentale Vito Volterra, Universit\`a degli Studi di Roma ``Tor Vergata''\\\texttt{accardi@volterra.uniroma2.it} \and Abdessatar Souissi\\ Department of mathematical physics,\\ Preparatory institute for engineering studies, University of Monastir, Tunisia\\\texttt{a.souaissi@qu.edu.sa}}

\maketitle

\begin{abstract}
A Markov operator on a matrix algebra, that is, a unital completely positive map, is called positivity improving if it maps every nonzero positive element to an invertible one. This is the noncommutative counterpart of a stochastic matrix with strictly positive entries. We show that positivity improvement is self-dual and equivalent to a faithful quantum Markov--Dobrushin--Doeblin minorization. We also compare the scalar trace minorization with the optimal common-lower-bound construction of the quantum Markov--Dobrushin coefficient.  We then characterize positivity improvement through the Kraus space of the map: a Markov operator is positivity improving if and only if the Hilbert--Schmidt orthogonal complement of its Kraus space contains no operator of rank one. As a consequence the Kraus rank of a positivity improving Markov operator on the algebra of $d\times d$ matrices is at least $2d-1$, and an explicit Hankel-type channel shows that this bound is attained in every dimension. Although deciding positivity improvement is NP-hard in general, we obtain complete and effectively checkable classifications for qubits and qutrits, together with families of examples (Pauli, Weyl and Werner--Holevo channels) showing that all our estimates are optimal. Finally we show that, in infinite dimensions, positivity improvement no longer implies any contraction of the Markov operator.
\end{abstract}

\par

\par

\section{Introduction}\label{sec:intro}

Let $p=(p(x,y))_{x,y}$ be a stochastic matrix on a finite set $S$. If every entry of $p$ is strictly positive, then $p$ sends every nonzero nonnegative vector to a strictly positive one, and the Perron--Frobenius theorem yields a unique strictly positive invariant distribution together with geometric convergence towards it. The rate of convergence is governed by the \emph{Dobrushin ergodicity coefficient}
\[
\delta(p)=\tfrac12\max_{x,x'}\sum_{y}|p(x,y)-p(x',y)|,
\]
introduced by Dobrushin \cite{Dob56} in his study of  Markov chains. The basic estimate behind this theory is the \emph{Markov--Dobrushin-Doeblin  minorization}: if $p(x,\cdot)\ge \kappa\,\nu(\cdot)$ for all $x$ and some probability measure $\nu$, then $\delta(p)\le 1-\kappa$; see Seneta \cite{Sen06}. Operators that send nonzero positive elements to strictly positive ones are called \emph{positivity improving}. In infinite dimensions this notion is at the heart of the uniqueness theory of ground states in quantum field theory and statistical mechanics, through the work of Gross \cite{Gro72}, Faris \cite{Far72}, and Reed--Simon \cite{RS78}.

In quantum probability the role of stochastic matrices is played by \emph{quantum Markov operators}, i.e.\ unital completely positive maps on a $C^*$-algebra, which are the transition operators of quantum Markov chains in the sense of Accardi \cite{Acc75,AF83}, and the transfer operators of finitely correlated (matrix product) states \cite{FNW92}. The noncommutative Perron--Frobenius theory of such maps was initiated by Evans and H{\o}egh-Krohn \cite{EHK78}, who introduced irreducibility for positive maps on $C^*$-algebras, and developed further by Farenick \cite{Far96}. Sanz, P\'erez-Garc\'{\i}a, Wolf and Cirac \cite{SPWC10} studied \emph{primitive} maps, i.e.\ maps some power of which is \emph{strictly positive} (in our terminology, positivity improving), and proved a quantum version of Wielandt's inequality; see also Wolf's lecture notes \cite{Wol12}. On the ergodic side, Dobrushin's coefficient was extended to positive contractions on preduals of von Neumann algebras by Mukhamedov \cite{Muk13}. A quantum Markov--Dobrushin inequality was established by Accardi, Lu and Souissi \cite{AccLuSou21} and subsequently refined in the study of inhomogeneous quantum processes \cite{S26}. Related contraction techniques are relevant to exponential mixing and matrix product states \cite{SB25}; see also the ergodic theory of quantum processes \cite{MS11}. Dobrushin coefficients for Markov operators on general cones were investigated by Gaubert and Qu \cite{GQ15}. Contraction properties of quantum channels were also studied through the trace norm by Ruskai \cite{Rus94} and Raginsky \cite{Rag02}, and through Hilbert's projective metric by Reeb, Kastoryano and Wolf \cite{RKW11}. Quantum Markov chains and their ergodic properties on lattices and trees have been studied from the operator-algebraic viewpoint in \cite{ASS20,MS19}.

Two features make positivity improvement in the quantum setting much more subtle than in the classical one. First, positivity of a matrix in ${M_d(\mathbb{C})}$ is not an entrywise property, so there is no ``positive entries'' criterion. Second, Gaubert and Qu \cite{GQ17} proved that \emph{deciding whether a Kraus map is strictly positive is NP-hard}. Hence one cannot expect simple criteria in arbitrary dimension, and it becomes natural to ask for (a) structural characterizations that explain where the difficulty lies, (b) sharp necessary conditions, and (c) complete and effectively checkable classifications in the physically most relevant low dimensions.

The purpose of this paper is to answer these questions. Our main results are as follows.

\begin{itemize}
\item \emph{Self-duality and the Markov--Dobrushin condition} (Proposition~\ref{prop:dual} and Theorem~\ref{thm:MD}). A Markov operator $P$ on ${\mathcal{B}(\mathcal{H})}$, $\dim{\mathcal{H}}=d<\infty$, is positivity improving if and only if its predual (the corresponding quantum channel) is, and if and only if it satisfies the quantum Markov--Dobrushin condition
\[
P(a)\ \ge\ \kappa\,{\operatorname{Tr}}(\sigma a)\,{\mathbf{1}},\qquad a\ge 0,
\]
for some $\kappa>0$ and some faithful state $\sigma$. The optimal constant for $\sigma={\mathbf{1}}/d$ is $\kappa=d\,\lambda(P)$, where $\lambda(P)$ is the smallest eigenvalue of an output of a pure state.

\item \emph{Ergodic consequences} (Theorem~\ref{thm:ergodic}). If $P$ is positivity improving, then the quantum Dobrushin coefficient satisfies $\delta(P)\le 1-d\lambda(P)<1$, $P$ has a unique invariant state, this state is faithful, $1$ is an algebraically simple eigenvalue, and every other eigenvalue $\mu$ satisfies $|\mu|\le 1-d\lambda(P)$. All these bounds are attained (Proposition~\ref{prop:pauli} and Example~\ref{ex:WH}).

\item \emph{Kraus-space characterization} (Theorem~\ref{thm:kraus}). $P$ is positivity improving if and only if the Hilbert--Schmidt orthogonal complement ${\mathcal{K}}(P)^{\perp}$ of its Kraus space contains no rank-one operator. Equivalently, the kernel of the Choi matrix of $P$ is a \emph{completely entangled subspace} in the sense of Parthasarathy \cite{Par04}. This links the problem to the geometry of the Segre variety and explains its computational hardness.

\item \emph{Sharp Kraus-rank bound} (Corollary~\ref{cor:rank} and Proposition~\ref{prop:hankel}). The Kraus rank of a positivity improving Markov operator on ${M_d(\mathbb{C})}$ is at least $2d-1$. This bound is attained in every dimension by an explicit bistochastic ``Hankel channel''.

\item \emph{Classification for $d=2$ and $d=3$} (Theorems~\ref{thm:d2} and \ref{thm:d3}). For qubits, $P$ is positivity improving if and only if either its Choi matrix is invertible, or its Kraus rank equals $3$ and the generator of ${\mathcal{K}}(P)^\perp$ is invertible. For qutrits we obtain an analogous classification in terms of the adjugates of the elements of ${\mathcal{K}}(P)^\perp$, which reduces the problem to the absence of common zeros of at most nine quadratic forms in at most four variables.

\item \emph{Infinite dimensions} (Section~\ref{sec:infinite}). The Markov--Dobrushin estimate survives, but positivity improvement no longer implies any contraction: we construct a positivity improving normal channel with $\delta=1$.
\end{itemize}

We emphasize that several natural guesses are false. Invertibility of all Kraus operators is neither necessary nor sufficient for positivity improvement in any dimension (Remark~\ref{rem:invertible}). Moreover, strict positivity of the images of the rank-one projections of a single orthonormal basis does not suffice (Example~\ref{ex:basis}).

The paper is organized as follows. Section~\ref{sec:prelim} fixes notation and proves the basic structural properties of positivity improving maps. Section~\ref{sec:MD} contains the Markov--Dobrushin theory and its ergodic consequences. Section~\ref{sec:kraus} is devoted to the Kraus-space and Choi-matrix characterization and to the sharp rank bound. Sections~\ref{sec:d2} and \ref{sec:d3} give the classifications in dimensions two and three, and Section~\ref{sec:infinite} discusses the infinite-dimensional situation.

\section{Preliminaries and Basic Properties}\label{sec:prelim}

Throughout, except in Section~\ref{sec:infinite}, ${\mathcal{H}}$ is a complex Hilbert space of finite dimension $d\ge 2$, with inner product ${\langle \cdot, \cdot\rangle}$ linear in the second variable, and ${\mathcal{B}(\mathcal{H})}\cong{M_d(\mathbb{C})}$ is the algebra of all operators on ${\mathcal{H}}$. We write $a\ge 0$ for positive semidefinite operators and $a>0$ for positive \emph{invertible} ones. The set of density matrices (states) is ${\mathfrak{S}(\mathcal{H})}=\{\rho\ge0:{\operatorname{Tr}}\rho=1\}$, and for a unit vector $\psi$ we put $e_\psi:={|\psi\rangle\langle \psi|}$. We use the trace norm $\|x\|_1={\operatorname{Tr}}|x|$, the operator norm $\|a\|$, and the Hilbert--Schmidt inner product ${\langle x, y\rangle}_2={\operatorname{Tr}}(x^*y)$. For a self-adjoint $a$ we denote by ${\lambda_{\min}}(a)$ and $\lambda_{\max}(a)$ its extreme eigenvalues and put ${\operatorname{osc}}(a):=\lambda_{\max}(a)-{\lambda_{\min}}(a)$. We fix an orthonormal basis $(f_j)_{j=1}^d$ of ${\mathcal{H}}$, write $E_{jk}={|f_j\rangle\langle f_k|}$, and for $\psi=\sum_j\psi_jf_j$ we let $\bar\psi:=\sum_j\overline{\psi_j}f_j$. The transpose $x^{T}$ is taken with respect to $(f_j)$.

\begin{definition}\label{def:markov}
A \emph{quantum Markov operator} on ${\mathcal{B}(\mathcal{H})}$ is a linear map $P:{\mathcal{B}(\mathcal{H})}\to{\mathcal{B}(\mathcal{H})}$ that is completely positive and unital, $P({\mathbf{1}})={\mathbf{1}}$. Its \emph{predual} ${P_{*}}:{\mathcal{B}(\mathcal{H})}\to{\mathcal{B}(\mathcal{H})}$ is defined by
\[
{\operatorname{Tr}}\big({P_{*}}(x)\,a\big)={\operatorname{Tr}}\big(x\,P(a)\big),\qquad x,a\in{\mathcal{B}(\mathcal{H})} .
\]
\end{definition}

Thus $P$ describes the evolution of observables (Heisenberg picture), and ${P_{*}}$ is a completely positive trace-preserving map, i.e.\ a quantum channel (Schr\"odinger picture). By the theorems of Stinespring \cite{Sti55}, Kraus \cite{Kra83} and Choi \cite{Cho75}, there are $K_1,\dots,K_r\in{\mathcal{B}(\mathcal{H})}$ such that
\begin{equation}\label{eq:kraus}
{P_{*}}(x)=\sum_{i=1}^rK_ixK_i^*,\qquad P(a)=\sum_{i=1}^rK_i^*aK_i,\qquad \sum_{i=1}^rK_i^*K_i={\mathbf{1}} .
\end{equation}
The \emph{Kraus space} ${\mathcal{K}}(P):=\operatorname{span}\{K_1,\dots,K_r\}\subset{\mathcal{B}(\mathcal{H})}$ does not depend on the chosen Kraus decomposition, since two decompositions are related by an isometric matrix \cite{Cho75,Wol12}. Its dimension is the \emph{Kraus rank} $r(P)$, which is the minimal number of Kraus operators. The \emph{Choi matrix} of $P$ is
\begin{equation}\label{eq:choi}
C_P:=\sum_{j,k=1}^dE_{jk}\otimes{P_{*}}(E_{jk})\in\mathcal{B}({\mathcal{H}}\otimes{\mathcal{H}}).
\end{equation}
It is positive, and ${\operatorname{rank}} C_P=r(P)$ (Lemma~\ref{lem:choi} below).

\begin{definition}\label{def:PI}
A positive linear map $T:{\mathcal{B}(\mathcal{H})}\to{\mathcal{B}(\mathcal{H})}$ is \emph{positivity improving} if
\[
a\ge0,\ a\neq0\quad\Longrightarrow\quad T(a)>0 ,
\]
that is, $T(a)$ is invertible, or equivalently its support projection is ${\mathbf{1}}$. A Markov operator $P$ is positivity improving if $P$ is. (By Proposition~\ref{prop:dual} this holds if and only if ${P_{*}}$ is positivity improving.)
\end{definition}

In the literature on Perron--Frobenius theory for cones, such maps are also called \emph{strictly positive} \cite{SPWC10,GQ17}. For a positive map $T$ we write $T^\sharp$ for its adjoint with respect to the trace pairing, ${\operatorname{Tr}}(T^\sharp(x)a)={\operatorname{Tr}}(xT(a))$. In particular $P^\sharp={P_{*}}$ and $({P_{*}})^\sharp=P$.

\begin{proposition}[Pure-state criterion and self-duality]\label{prop:dual}
For a positive linear map $T$ on ${\mathcal{B}(\mathcal{H})}$ the following are equivalent:
\begin{enumerate}[label=(\roman*)]
\item $T$ is positivity improving;
\item $T(e_\psi)>0$ for every unit vector $\psi\in{\mathcal{H}}$;
\item ${\langle \varphi, T(e_\psi)\varphi\rangle}>0$ for all unit vectors $\varphi,\psi\in{\mathcal{H}}$;
\item $T^\sharp$ is positivity improving.
\end{enumerate}
In particular, a Markov operator $P$ is positivity improving if and only if the quantum channel ${P_{*}}$ is positivity improving.
\end{proposition}

\begin{proof}
(i)$\Rightarrow$(ii) is obvious. (ii)$\Leftrightarrow$(iii) holds because a positive semidefinite operator $b$ is invertible if and only if ${\langle \varphi, b\varphi\rangle}>0$ for all $\varphi\neq0$.
(iii)$\Rightarrow$(i): let $a\ge0$, $a\ne0$, and let $\psi$ be a unit eigenvector of $a$ for the eigenvalue $\|a\|>0$. Then $a\ge\|a\|\,e_\psi$, so positivity of $T$ gives $T(a)\ge\|a\|\,T(e_\psi)$. Since (iii) implies (ii), the right-hand side is invertible, hence so is $T(a)$.
(iii)$\Leftrightarrow$(iv): for unit vectors $\varphi,\psi$,
\[
{\langle \varphi, T(e_\psi)\varphi\rangle}={\operatorname{Tr}}\big(e_\varphi T(e_\psi)\big)={\operatorname{Tr}}\big(T^\sharp(e_\varphi)e_\psi\big)={\langle \psi, T^\sharp(e_\varphi)\psi\rangle}.
\]
Hence condition (iii) for $T$ is the same as condition (iii) for $T^\sharp$, with the roles of $\varphi$ and $\psi$ exchanged. Applying (i)$\Leftrightarrow$(iii) to $T^\sharp$ concludes the proof.
\end{proof}

\begin{proposition}[Structural properties]\label{prop:struct}
Let $P$ be a positivity improving Markov operator on ${\mathcal{B}(\mathcal{H})}$.
\begin{enumerate}[label=(\roman*)]
\item For every Markov operator $Q$ on ${\mathcal{B}(\mathcal{H})}$, the composition $Q\circ P$ is positivity improving. In particular, $P^n$ is positivity improving for all $n\ge1$.
\item For every Markov operator $Q$ and every $t\in[0,1)$, the operator $(1-t)P+tQ$ is positivity improving. In particular, the positivity improving Markov operators form a convex set.
\item ${P_{*}}$ is irreducible in the sense of Evans and H{\o}egh-Krohn \cite{EHK78}: there is no projection $p\notin\{0,{\mathbf{1}}\}$ with ${P_{*}}(p{\mathcal{B}(\mathcal{H})} p)\subseteq p{\mathcal{B}(\mathcal{H})} p$. Moreover ${P_{*}}$ is primitive in the sense of \cite{SPWC10}.
\item The set of positivity improving Markov operators is relatively open in the set of all Markov operators on ${\mathcal{B}(\mathcal{H})}$.
\end{enumerate}
\end{proposition}

\begin{proof}
(i) If $a\ge0$, $a\ne0$, then $P(a)\ge c{\mathbf{1}}$ for some $c>0$. Since $Q$ is positive and unital, $Q(P(a))\ge cQ({\mathbf{1}})=c{\mathbf{1}}$.
(ii) $(1-t)P(a)+tQ(a)\ge(1-t)P(a)>0$.
(iii) If $p\ne0$ is a projection with ${P_{*}}(p{\mathcal{B}(\mathcal{H})} p)\subseteq p{\mathcal{B}(\mathcal{H})} p$, then ${P_{*}}(p)\le\|{P_{*}}(p)\|\,p$. Since ${P_{*}}(p)$ is invertible by Proposition~\ref{prop:dual}, this forces $p={\mathbf{1}}$. Primitivity, i.e.\ positivity improvement of some power of ${P_{*}}$, holds trivially with the first power.
(iv) The map $\psi\mapsto{\lambda_{\min}}({P_{*}}(e_\psi))$ depends continuously on the pair $(P,\psi)$ by Weyl's inequality $|{\lambda_{\min}}(b)-{\lambda_{\min}}(c)|\le\|b-c\|$. By Proposition~\ref{prop:dual} and compactness of the unit sphere, $P$ is positivity improving if and only if $\lambda(P):=\min_{\|\psi\|=1}{\lambda_{\min}}({P_{*}}(e_\psi))>0$. This quantity depends continuously on $P$.
\end{proof}

\begin{remark}\label{rem:invertible}
Invertibility of all Kraus operators is neither sufficient nor necessary for positivity improvement, in every dimension $d\ge2$. If $U$ is unitary, then $P(a)=U^*aU$ has the single invertible Kraus operator $U$, yet ${P_{*}}(e_\psi)=e_{U\psi}$ has rank one. Conversely, the completely depolarizing channel ${P_{*}}(x)={\operatorname{Tr}}(x){\mathbf{1}}/d$ admits the Kraus decomposition $K_{jk}=d^{-1/2}E_{jk}$, $1\le j,k\le d$, into rank-one operators, and it is clearly positivity improving.
\end{remark}

\begin{example}[Classical Markov operators]\label{ex:classical}
Let $p$ be a $d\times d$ stochastic matrix and let ${P_{*}}(x)=\sum_{x',y}p(x',y){\langle f_{x'}, xf_{x'}\rangle}E_{yy}$ be the associated measure-and-prepare channel, with Kraus operators $\sqrt{p(x',y)}\,{|f_y\rangle\langle f_{x'}|}$. Then ${P_{*}}(e_\psi)=\sum_y\big(\sum_{x'}|\psi_{x'}|^2p(x',y)\big)E_{yy}$. Hence ${P_{*}}$ is positivity improving if and only if $p(x',y)>0$ for all $x',y$, that is, if and only if $p$ is positivity improving in the classical sense.
\end{example}

\section{The Quantum Markov--Dobrushin Condition}\label{sec:MD}

\subsection{The quantum Dobrushin coefficient}

Following \cite{Muk13,GQ15,Rus94,Rag02}, the \emph{Dobrushin ergodicity coefficient} of a Markov operator $P$ on ${\mathcal{B}(\mathcal{H})}$ is
\begin{equation}\label{eq:dob}
\delta(P):=\sup_{\rho,\sigma\in{\mathfrak{S}(\mathcal{H})}}\tfrac12\,\|{P_{*}}(\rho)-{P_{*}}(\sigma)\|_1\in[0,1].
\end{equation}
For the channel of Example~\ref{ex:classical} restricted to diagonal states, \eqref{eq:dob} is exactly Dobrushin's classical coefficient of $p$.

\begin{lemma}\label{lem:dob}
Let $P,Q$ be Markov operators on ${\mathcal{B}(\mathcal{H})}$ and let $\mathcal{T}_0:=\{X=X^*\in{\mathcal{B}(\mathcal{H})}:{\operatorname{Tr}} X=0\}$.
\begin{enumerate}[label=(\roman*)]
\item $\delta(P)=\sup\big\{\|{P_{*}}(X)\|_1/\|X\|_1:\ X\in\mathcal{T}_0,\ X\ne0\big\}$.
\item $\delta(P\circ Q)\le\delta(P)\,\delta(Q)$.
\item (Heisenberg form) $\delta(P)=\sup\big\{{\operatorname{osc}}(P(a)):\ a=a^*,\ {\operatorname{osc}}(a)\le1\big\}$.
\end{enumerate}
\end{lemma}

\begin{proof}
(i) Let $X\in\mathcal{T}_0\setminus\{0\}$ have Jordan decomposition $X=X_+-X_-$. Then ${\operatorname{Tr}} X_+={\operatorname{Tr}} X_-=\|X\|_1/2=:t>0$, and $X=t(\rho-\sigma)$ with $\rho=X_+/t$, $\sigma=X_-/t\in{\mathfrak{S}(\mathcal{H})}$. Hence $\|{P_{*}}(X)\|_1/\|X\|_1=\tfrac12\|{P_{*}}(\rho)-{P_{*}}(\sigma)\|_1$. Conversely, $\rho-\sigma\in\mathcal{T}_0$ and $\|\rho-\sigma\|_1\le2$. Together these give (i).
(ii) Since $(P\circ Q)_*={Q_{*}}\circ{P_{*}}$ and ${P_{*}}$ leaves $\mathcal{T}_0$ invariant (being Hermiticity- and trace-preserving), (ii) follows from (i).
(iii) For $Y\in\mathcal{T}_0$ and $a=a^*$, writing $Y=t(\rho-\sigma)$ as above,
\begin{equation}\label{eq:osc}
|{\operatorname{Tr}}(Ya)|=t\,|{\operatorname{Tr}}(\rho a)-{\operatorname{Tr}}(\sigma a)|\le\tfrac12\|Y\|_1{\operatorname{osc}}(a),
\end{equation}
with equality when $a$ is the spectral projection of $Y$ onto $(0,\infty)$, for then ${\operatorname{osc}}(a)\le1$ and ${\operatorname{Tr}}(Ya)={\operatorname{Tr}} Y_+=\tfrac12\|Y\|_1$. Since ${\operatorname{osc}}(b)=\sup_{\rho,\sigma\in{\mathfrak{S}(\mathcal{H})}}{\operatorname{Tr}}((\rho-\sigma)b)$ for $b=b^*$, we get
\[
\sup_{{\operatorname{osc}}(a)\le1}{\operatorname{osc}}(P(a))=\sup_{\rho,\sigma}\ \sup_{{\operatorname{osc}}(a)\le1}{\operatorname{Tr}}\big(({P_{*}}(\rho)-{P_{*}}(\sigma))a\big)=\sup_{\rho,\sigma}\tfrac12\|{P_{*}}(\rho)-{P_{*}}(\sigma)\|_1 .
\]
\end{proof}

Here ${Q_{*}}$ denotes the predual of $Q$. Item (iii) is the exact noncommutative analogue of the classical formula $\delta(p)=\sup\{{\operatorname{osc}}(pf):{\operatorname{osc}}(f)\le1\}$; compare \cite{GQ15}.

\subsection{Optimal common lower bounds and the quantum Markov--Dobrushin inequality}\label{subsec:optimalMD}

The quantum Markov--Dobrushin inequality was introduced by Accardi, Lu and Souissi
\cite{AccLuSou21}; a maximal-trace common-lower-bound formulation was subsequently
used in the study of inhomogeneous quantum processes \cite{S26}. Related mixing
conditions and applications to matrix product states appear in \cite{SB25}.
We formulate the variational coefficient carefully, distinguishing an optimal
minorization bound from the actual trace-distance contraction coefficient.

Let
\[
\mathcal P_1({\mathcal{H}}):=\{{|\xi\rangle\langle \xi|}:\xi\in{\mathcal{H}},\ \|\xi\|=1\}.
\]
For a quantum channel $\Phi:{\mathcal{B}(\mathcal{H})}\to{\mathcal{B}(\mathcal{H})}$, define the compact convex set
\begin{equation}\label{eq:commonMD}
\mathcal L_{\rm MD}(\Phi):=\{B\in{\mathcal{B}(\mathcal{H})}:B\ge0,\ B\le\Phi(E)
\text{ for every }E\in\mathcal P_1({\mathcal{H}})\}.
\end{equation}
\begin{lemma}[Existence of maximal-trace common lower bounds]\label{Lem_MD}
The set $\mathcal L_{\rm MD}(\Phi)$ is nonempty and compact. Consequently
$\max\{{\operatorname{Tr}} B:B\in\mathcal L_{\rm MD}(\Phi)\}$ exists and lies in $[0,1]$.
\end{lemma}
\begin{proof}
The zero operator is admissible. The set is closed, and every admissible $B$
satisfies $0\le B\le\Phi(E_0)$ for any fixed rank-one projection $E_0$.
It is therefore bounded and, in finite dimension, compact. Taking traces
shows $0\le{\operatorname{Tr}} B\le{\operatorname{Tr}}\Phi(E_0)=1$.
\end{proof}

\begin{definition}[Markov--Dobrushin infimum set]\label{def:optimalMD}
The \emph{Markov--Dobrushin infimum set} is the set of maximal-trace common
lower bounds
\begin{equation}\label{infMD}
\begin{split}
\mathcal I^{{\operatorname{Tr}}}_{\rm MD}(\Phi):=\{\kappa_\Phi\in{\mathcal{B}(\mathcal{H})}:\;&
0\le\kappa_\Phi\le\Phi(E)\quad\forall E\in\mathcal P_1({\mathcal{H}}),\\
&{\operatorname{Tr}}\kappa_\Phi\ge{\operatorname{Tr}} B\quad\forall B\in\mathcal L_{\rm MD}(\Phi)\}.
\end{split}
\end{equation}
For any $\kappa_\Phi\in\mathcal I^{{\operatorname{Tr}}}_{\rm MD}(\Phi)$, put
\begin{equation}\label{eq:etaMD}
\eta_{\rm MD}(\Phi):=1-{\operatorname{Tr}}\kappa_\Phi.
\end{equation}
This number is independent of the choice of the maximizing operator.
\end{definition}

The construction avoids assuming that the family $\{\Phi(E):E\in
\mathcal P_1({\mathcal{H}})\}$ has a greatest lower bound in the L\"owner order.
It optimizes the trace over all common positive lower bounds, not the
trace-distance contraction coefficient itself.

\begin{theorem}[Quantum Markov--Dobrushin inequality]\label{thm:optimalMD}
For all states $\rho,\sigma\in{\mathfrak{S}(\mathcal{H})}$,
\begin{equation}\label{MDineq}
\|\Phi(\rho)-\Phi(\sigma)\|_{\rm TV}
\le\eta_{\rm MD}(\Phi)\|\rho-\sigma\|_{\rm TV},
\qquad\|X\|_{\rm TV}:=\tfrac12\|X\|_1
\quad (X=X^*,\ {\operatorname{Tr}} X=0).
\end{equation}
More precisely, the optimal trace-norm contraction coefficient
$\delta(\Phi):=\sup_{\rho\ne\sigma}\|\Phi(\rho)-\Phi(\sigma)\|_1/
\|\rho-\sigma\|_1$ satisfies
\begin{equation}\label{eq:deltaeta}
\delta(\Phi)\le\eta_{\rm MD}(\Phi)\le1.
\end{equation}
\end{theorem}
\begin{proof}
Choose $\kappa_\Phi\in\mathcal I^{{\operatorname{Tr}}}_{\rm MD}(\Phi)$ and put
$c={\operatorname{Tr}}\kappa_\Phi$. Since every state is a convex combination of rank-one
projections, $\Phi(\rho)\ge\kappa_\Phi$ for every state $\rho$.
If $c=1$, positivity and equality of traces force
$\Phi(\rho)=\kappa_\Phi$ for every state, so $\delta(\Phi)=0$.
If $c<1$, define
\[
R(X)=\frac{\Phi(X)-{\operatorname{Tr}}(X)\kappa_\Phi}{1-c}.
\]
For positive $X$, the spectral decomposition of $X$ shows $R(X)\ge0$;
moreover $R$ preserves the trace. For traceless self-adjoint $X$,
$\Phi(X)=(1-c)R(X)$, and positivity and trace preservation give
$\|R(X)\|_1\le\|X\|_1$. Thus $\delta(\Phi)\le1-c$.
\end{proof}

\begin{remark}[Optimality and its limitation]\label{rem:etaoptimal}
The value $\eta_{\rm MD}(\Phi)$ is the smallest contraction bound
obtainable by subtracting a \emph{single common positive lower bound} from
all outputs. It need not equal $\delta(\Phi)$: for a classical
measure-and-prepare channel with output distributions
$p_1=(1/2,1/2,0)$, $p_2=(1/2,0,1/2)$, and
$p_3=(0,1/2,1/2)$, the only common nonnegative lower bound is zero,
so $\eta_{\rm MD}=1$, whereas the classical Dobrushin coefficient is $1/2$.
In particular, calling $\eta_{\rm MD}$ the \emph{minimal contraction
factor} without restricting to common-lower-bound estimates would be
incorrect.
\end{remark}

\begin{proposition}[Comparison with scalar minorization]\label{prop:MDcompare}
For $\Phi={P_{*}}$ and $\lambda(P)$ as in \eqref{eq:lambda},
\[
\delta(P)\le\eta_{\rm MD}({P_{*}})\le1-d\lambda(P).
\]
If $P$ is positivity improving, then $\eta_{\rm MD}({P_{*}})<1$.
\end{proposition}
\begin{proof}
Theorem~\ref{thm:optimalMD} gives the first inequality. By definition
${P_{*}}(e_\psi)\ge\lambda(P){\mathbf{1}}$ for every unit $\psi$, so
$\lambda(P){\mathbf{1}}\in\mathcal L_{\rm MD}({P_{*}})$; maximality of the trace
gives the second inequality. If $P$ is positivity improving, then $\lambda(P)>0$, so $\eta_{\rm MD}({P_{*}})<1$.
\end{proof}

\subsection{Markov--Dobrushin condition}

\begin{definition}\label{def:MD}
Let $\kappa\in(0,1]$ and $\sigma\in{\mathfrak{S}(\mathcal{H})}$. A Markov operator $P$ satisfies the \emph{Markov--Dobrushin condition} $\mathrm{MD}(\kappa,\sigma)$ if
\begin{equation}\label{eq:MD}
P(a)\ \ge\ \kappa\,{\operatorname{Tr}}(\sigma a)\,{\mathbf{1}}\qquad\text{for all }a\ge0,
\end{equation}
or, equivalently (by duality, testing \eqref{eq:MD} against states), if ${P_{*}}(\rho)\ge\kappa\,\sigma$ for all $\rho\in{\mathfrak{S}(\mathcal{H})}$.
\end{definition}

For the classical channel of Example~\ref{ex:classical}, $\mathrm{MD}(\kappa,\sigma)$ with diagonal $\sigma=\sum_y\nu(y)E_{yy}$ is exactly the Doeblin minorization $p(x,\cdot)\ge\kappa\,\nu(\cdot)$.

\begin{theorem}[Markov--Dobrushin inequality]\label{thm:MDineq}
If $P$ satisfies $\mathrm{MD}(\kappa,\sigma)$, then $\delta(P)\le1-\kappa$.
\end{theorem}

\begin{proof}
If $\kappa=1$, then ${P_{*}}(\rho)-\sigma$ is positive with trace zero, hence ${P_{*}}(\rho)=\sigma$ for all $\rho$, and $\delta(P)=0$. Let $\kappa<1$ and define
\[
R(x):=\frac{1}{1-\kappa}\big({P_{*}}(x)-\kappa\,{\operatorname{Tr}}(x)\,\sigma\big),\qquad x\in{\mathcal{B}(\mathcal{H})} .
\]
By $\mathrm{MD}(\kappa,\sigma)$ the map $R$ is positive, and it is trace preserving. A positive trace-preserving map contracts the trace norm on self-adjoint elements: $\|R(X)\|_1\le{\operatorname{Tr}} R(X_+)+{\operatorname{Tr}} R(X_-)=\|X\|_1$. For $X\in\mathcal{T}_0$ we have ${P_{*}}(X)=(1-\kappa)R(X)$, so $\|{P_{*}}(X)\|_1\le(1-\kappa)\|X\|_1$, and Lemma~\ref{lem:dob}(i) concludes the proof.
\end{proof}

The next result, a corrected and sharpened form of the ``strict positivity on rank-one projections'' principle, identifies positivity improvement with the Markov--Dobrushin condition. Define
\begin{equation}\label{eq:lambda}
\lambda(P):=\min_{\psi\in{\mathcal{H}},\ \|\psi\|=1}{\lambda_{\min}}\big({P_{*}}(e_\psi)\big).
\end{equation}
The minimum exists by continuity and compactness (see the proof of Proposition~\ref{prop:struct}(iv)), and $0\le d\,\lambda(P)\le{\operatorname{Tr}}{P_{*}}(e_\psi)=1$.

\begin{theorem}[Positivity improvement $=$ Markov--Dobrushin]\label{thm:MD}
For a Markov operator $P$ on ${\mathcal{B}(\mathcal{H})}$ the following are equivalent:
\begin{enumerate}[label=(\roman*)]
\item $P$ is positivity improving;
\item $P$ satisfies $\mathrm{MD}(\kappa,\sigma)$ for some $\kappa>0$ and some faithful state $\sigma$;
\item $\lambda(P)>0$.
\end{enumerate}
In this case $P$ satisfies $\mathrm{MD}(d\lambda(P),{\mathbf{1}}/d)$, and $d\lambda(P)$ is the largest constant $\kappa$ for which $\mathrm{MD}(\kappa,{\mathbf{1}}/d)$ holds.
\end{theorem}

\begin{proof}
(i)$\Rightarrow$(iii) follows from Proposition~\ref{prop:dual} and the attainment of the minimum in \eqref{eq:lambda}.
(iii)$\Rightarrow$(ii): every $\rho\in{\mathfrak{S}(\mathcal{H})}$ has a spectral decomposition $\rho=\sum_kp_ke_{\psi_k}$ with $p_k\ge0$, $\sum_kp_k=1$. Hence ${P_{*}}(\rho)\ge\sum_kp_k\lambda(P){\mathbf{1}}=d\lambda(P)\cdot{\mathbf{1}}/d$, which is $\mathrm{MD}(d\lambda(P),{\mathbf{1}}/d)$.
(ii)$\Rightarrow$(i): ${P_{*}}(\rho)\ge\kappa\sigma>0$ for every state $\rho$, so ${P_{*}}$ is positivity improving, and so is $P$ by Proposition~\ref{prop:dual}.
Finally, if $\mathrm{MD}(\kappa,{\mathbf{1}}/d)$ holds, then taking $\rho=e_\psi$ gives ${\lambda_{\min}}({P_{*}}(e_\psi))\ge\kappa/d$ for all $\psi$, i.e.\ $\kappa\le d\lambda(P)$.
\end{proof}

The following lemma states precisely what a positivity condition on a \emph{finite} family of rank-one projections yields.

\begin{lemma}\label{lem:finite}
Let $F$ be a finite family of unit vectors such that ${P_{*}}(e_\psi)>0$ for all $\psi\in F$. Then $\lambda_F:=\min_{\psi\in F}{\lambda_{\min}}({P_{*}}(e_\psi))>0$, and ${P_{*}}(\rho)\ge\lambda_F{\mathbf{1}}$ for every $\rho$ in the convex hull of $\{e_\psi:\psi\in F\}$.
\end{lemma}

\begin{proof}
Each ${\lambda_{\min}}({P_{*}}(e_\psi))$ is strictly positive and $F$ is finite. The second assertion follows by convexity.
\end{proof}

The convex hull in Lemma~\ref{lem:finite} is a proper subset of ${\mathfrak{S}(\mathcal{H})}$. The next example shows that the hypothesis of Lemma~\ref{lem:finite}, even for an orthonormal basis, does not imply positivity improvement. Hence it does not imply $\delta(P)<1$ via Theorem~\ref{thm:MDineq}.

\begin{example}\label{ex:basis}
On ${\mathbb{C}}^2$ let ${|\pm\rangle}=(f_1\pm f_2)/\sqrt2$ and let ${P_{*}}(x)={\langle +|}x{|+\rangle}\tfrac12{\mathbf{1}}+{\langle -|}x{|-\rangle}e_{f_1}$, an entanglement-breaking channel. Then ${P_{*}}(e_{f_1})={P_{*}}(e_{f_2})=\tfrac14{\mathbf{1}}+\tfrac12e_{f_1}>0$, but ${P_{*}}(e_{-})=e_{f_1}$ is not invertible.
\end{example}

\subsection{Ergodic consequences}

\begin{theorem}\label{thm:ergodic}
Let $P$ be a positivity improving Markov operator on ${\mathcal{B}(\mathcal{H})}$ and put $\lambda=\lambda(P)$. Then:
\begin{enumerate}[label=(\roman*)]
\item $\delta(P)\le1-d\lambda<1$;
\item there is a unique $\rho_*\in{\mathfrak{S}(\mathcal{H})}$ with ${P_{*}}(\rho_*)=\rho_*$; it is faithful, and in fact $\rho_*\ge\lambda{\mathbf{1}}$;
\item for all $n\ge1$, all $\rho\in{\mathfrak{S}(\mathcal{H})}$ and all $a=a^*\in{\mathcal{B}(\mathcal{H})}$,
\[
\|{P_{*}}^n(\rho)-\rho_*\|_1\le2(1-d\lambda)^n,\qquad \big\|P^n(a)-{\operatorname{Tr}}(\rho_*a){\mathbf{1}}\big\|\le(1-d\lambda)^n{\operatorname{osc}}(a);
\]
\item $1$ is an algebraically simple eigenvalue of $P$ (equivalently of ${P_{*}}$), and every eigenvalue $\mu\ne1$ satisfies $|\mu|\le\delta(P)\le1-d\lambda$. In particular, the spectral gap
\[
\gamma(P):=1-\max\{|\mu|:\ \mu\in{\operatorname{spec}}(P)\setminus\{1\}\}
\]
satisfies $\gamma(P)\ge1-\delta(P)\ge d\lambda(P)>0$.
\end{enumerate}
\end{theorem}

\begin{proof}
(i) follows from Theorems~\ref{thm:MD} and \ref{thm:MDineq}.
(ii) By (i), ${P_{*}}$ is a strict contraction of the complete metric space $({\mathfrak{S}(\mathcal{H})},\|\cdot\|_1)$, so it has a unique fixed point by Banach's theorem. Moreover $\rho_*={P_{*}}(\rho_*)\ge\lambda{\mathbf{1}}$ by \eqref{eq:lambda} and the proof of Theorem~\ref{thm:MD}.
(iii) By Lemma~\ref{lem:dob}, $\|{P_{*}}^n(\rho)-\rho_*\|_1=\|{P_{*}}^n(\rho-\rho_*)\|_1\le\delta(P)^n\|\rho-\rho_*\|_1\le2\delta(P)^n$. For $a=a^*$ and $\rho\in{\mathfrak{S}(\mathcal{H})}$, by \eqref{eq:osc},
\[
\big|{\operatorname{Tr}}\big(\rho\,(P^n(a)-{\operatorname{Tr}}(\rho_*a){\mathbf{1}})\big)\big|=|{\operatorname{Tr}}(({P_{*}}^n(\rho)-\rho_*)a)|\le\tfrac12\|{P_{*}}^n(\rho)-\rho_*\|_1{\operatorname{osc}}(a)\le\delta(P)^n{\operatorname{osc}}(a).
\]
Taking the supremum over $\rho$ gives the operator-norm estimate.
(iv) $P$ and ${P_{*}}$ are transposes of each other with respect to the nondegenerate bilinear form $(x,a)\mapsto{\operatorname{Tr}}(xa)$, so they have the same eigenvalues with the same algebraic multiplicities. Let ${P_{*}}(X)=\mu X$ with $X\ne0$ and $\mu\ne1$. Since ${P_{*}}$ preserves the trace, $\mu{\operatorname{Tr}} X={\operatorname{Tr}} X$, so ${\operatorname{Tr}} X=0$. Write $X=H+iK$ with $H=(X+X^*)/2$ and $K=(X-X^*)/(2i)$ in $\mathcal{T}_0$. As ${P_{*}}$ preserves Hermiticity, Lemma~\ref{lem:dob} gives
\[
|\mu|^n\|X\|_1=\|{P_{*}}^n(X)\|_1\le\delta(P)^n(\|H\|_1+\|K\|_1)\le2\delta(P)^n\|X\|_1,
\]
whence $|\mu|\le2^{1/n}\delta(P)\to\delta(P)$. For simplicity, note that ${\mathcal{B}(\mathcal{H})}={\mathbb{C}}\rho_*\oplus\{X:{\operatorname{Tr}} X=0\}$ is a decomposition into ${P_{*}}$-invariant subspaces. By the same estimate, the spectral radius of ${P_{*}}$ restricted to the second summand is at most $\delta(P)<1$. Hence the eigenvalue $1$ has algebraic multiplicity one.
\end{proof}

\begin{remark}
(a) Theorem~\ref{thm:ergodic} provides the precise form of the relation between positivity improvement and spectral gaps: the gap is bounded below by $1-\delta(P)$, not by $\delta(P)$. All inequalities in Theorem~\ref{thm:ergodic}(i) and (iv) are equalities for the Pauli channels of Proposition~\ref{prop:pauli} and the Werner--Holevo channels of Example~\ref{ex:WH}.
(b) By the Birkhoff--Hopf theorem, positivity improving maps are also strict contractions for Hilbert's projective metric on the cone of positive definite matrices. This yields contraction rates for a wide family of distinguishability measures; see Reeb, Kastoryano and Wolf \cite{RKW11}.
(c) The quantity $\lambda(P)$ is the optimal value of a minimization of a biquadratic form over the product of two unit spheres, since $\lambda(P)=\min_{\varphi,\psi}{\langle \varphi, {P_{*}}(e_\psi)\varphi\rangle}$. This is the variational reason for the computational hardness established in \cite{GQ17}.
\end{remark}

\section{Kraus-Space and Choi-Matrix Characterization}\label{sec:kraus}

Recall that a linear subspace $\mathcal{S}\subseteq{\mathcal{H}}\otimes{\mathcal{H}}$ is \emph{completely entangled} if it contains no nonzero product vector $\xi\otimes\eta$ \cite{Par04,Bha06}.

\begin{lemma}\label{lem:choi}
Let $V:{\mathcal{B}(\mathcal{H})}\to{\mathcal{H}}\otimes{\mathcal{H}}$, $V(K):=\sum_{j=1}^df_j\otimes Kf_j$. Then:
\begin{enumerate}[label=(\roman*)]
\item $V$ is a unitary from $({\mathcal{B}(\mathcal{H})},{\langle \cdot, \cdot\rangle}_2)$ onto ${\mathcal{H}}\otimes{\mathcal{H}}$, and $V({|\varphi\rangle\langle \psi|})=\bar\psi\otimes\varphi$. In particular, $V$ maps the nonzero rank-one operators onto the nonzero product vectors.
\item $C_P=\sum_i{|V(K_i)\rangle\langle V(K_i)|}$ for any Kraus family $(K_i)$ of $P$. Hence $\operatorname{ran}C_P=V({\mathcal{K}}(P))$, $\ker C_P=V({\mathcal{K}}(P)^\perp)$, and ${\operatorname{rank}} C_P=r(P)$.
\item ${\langle \varphi, {P_{*}}(e_\psi)\varphi\rangle}=\sum_i|{\langle \varphi, K_i\psi\rangle}|^2={\langle \bar\psi\otimes\varphi, C_P(\bar\psi\otimes\varphi)\rangle}$ for all $\varphi,\psi\in{\mathcal{H}}$.
\end{enumerate}
\end{lemma}

\begin{proof}
(i) $V(E_{ab})=f_b\otimes f_a$, so $V$ maps an orthonormal basis onto an orthonormal basis. Moreover $V({|\varphi\rangle\langle \psi|})=\sum_j{\langle \psi, f_j\rangle}f_j\otimes\varphi=\bar\psi\otimes\varphi$.
(ii) $\sum_i{|V(K_i)\rangle\langle V(K_i)|}=\sum_{j,k}E_{jk}\otimes\sum_iK_iE_{jk}K_i^*=C_P$. The range of $\sum_i{|v_i\rangle\langle v_i|}$ is $\operatorname{span}\{v_i\}$, and $V$ is unitary.
(iii) The first equality follows from \eqref{eq:kraus}. For the second, ${\langle \bar\psi\otimes\varphi, V(K_i)\rangle}=\sum_j{\langle \bar\psi, f_j\rangle}{\langle \varphi, K_if_j\rangle}=\sum_j\psi_j{\langle \varphi, K_if_j\rangle}={\langle \varphi, K_i\psi\rangle}$, and we sum over $i$ using (ii).
\end{proof}

\begin{theorem}[Kraus-space characterization]\label{thm:kraus}
Let $P$ be a Markov operator on ${\mathcal{B}(\mathcal{H})}$ with Kraus family $(K_i)_{i=1}^r$. The following are equivalent:
\begin{enumerate}[label=(\roman*)]
\item $P$ is positivity improving;
\item for every $\psi\ne0$ the vectors $K_1\psi,\dots,K_r\psi$ span ${\mathcal{H}}$; equivalently, the bilinear system ${\langle \varphi, K_i\psi\rangle}=0$, $i=1,\dots,r$, has no solution with $\varphi\ne0\ne\psi$;
\item the Hilbert--Schmidt orthogonal complement ${\mathcal{K}}(P)^\perp$ contains no operator of rank one;
\item $\ker C_P$ is a completely entangled subspace of ${\mathcal{H}}\otimes{\mathcal{H}}$.
\end{enumerate}
In particular, positivity improvement depends only on the Kraus space ${\mathcal{K}}(P)$. If $P$ and $Q$ are Markov operators with ${\mathcal{K}}(P)\subseteq{\mathcal{K}}(Q)$ and $P$ is positivity improving, then so is $Q$.
\end{theorem}

\begin{proof}
By Proposition~\ref{prop:dual} and Lemma~\ref{lem:choi}(iii), $P$ is positivity improving if and only if $\sum_i|{\langle \varphi, K_i\psi\rangle}|^2>0$ for all $\varphi,\psi\neq0$, which is (ii). The first form of (ii) is equivalent to the second because $\varphi\perp K_i\psi$ for all $i$ exactly when $\varphi\perp\operatorname{span}\{K_i\psi\}$.
(ii)$\Leftrightarrow$(iii): ${\langle \varphi, K_i\psi\rangle}={\operatorname{Tr}}({|\psi\rangle\langle \varphi|}K_i)={\langle {|\varphi\rangle\langle \psi|}, K_i\rangle}_2$. Hence ${|\varphi\rangle\langle \psi|}\in{\mathcal{K}}(P)^\perp$ if and only if ${\langle \varphi, K_i\psi\rangle}=0$ for all $i$.
(iii)$\Leftrightarrow$(iv) follows from Lemma~\ref{lem:choi}(i),(ii). The last assertion follows from (iii), since ${\mathcal{K}}(Q)^\perp\subseteq{\mathcal{K}}(P)^\perp$.
\end{proof}

Condition (ii) is the formulation in terms of bilinear systems used by Gaubert and Qu \cite{GQ17}. Condition (iv) connects positivity improvement with the theory of completely entangled subspaces initiated by Parthasarathy \cite{Par04}; see also \cite{Wal02,Bha06,CMW08}. Geometrically, (iii) says that the projectivization of ${\mathcal{K}}(P)^\perp$ does not meet the Segre variety $\mathbb{P}^{d-1}\times\mathbb{P}^{d-1}\subset\mathbb{P}({\mathcal{B}(\mathcal{H})})$ of rank-one operators.

\begin{corollary}[Sharp Kraus-rank bound]\label{cor:rank}
Let $P$ be a Markov operator on ${M_d(\mathbb{C})}$.
\begin{enumerate}[label=(\roman*)]
\item If $P$ is positivity improving, then $r(P)\ge2d-1$.
\item If $r(P)=d^2$, i.e.\ $C_P$ is invertible, then $P$ is positivity improving.
\item For each $r\in\{2d-1,\dots,d^2\}$ there exist positivity improving Markov operators of Kraus rank $r$.
\end{enumerate}
\end{corollary}

\begin{proof}
(i) By Theorem~\ref{thm:kraus}, $\ker C_P$ is completely entangled. By Parthasarathy's theorem \cite{Par04} (see also \cite{Wal02,Bha06}), a completely entangled subspace of ${\mathbb{C}}^d\otimes{\mathbb{C}}^d$ has dimension at most $(d-1)^2$. Hence $d^2-r(P)\le(d-1)^2$.
(ii) Here ${\mathcal{K}}(P)^\perp=\{0\}$.
(iii) By Proposition~\ref{prop:hankel} below there is a positivity improving $H_d$ with $r(H_d)=2d-1$. Let $W_{ab}=X^aZ^b$ ($a,b\in\mathbb{Z}_d$) be the Weyl unitaries, which form an orthogonal basis of ${M_d(\mathbb{C})}$. Given $r$ with $2d-1\le r\le d^2$, choose $W_{a_1b_1},\dots,W_{a_mb_m}$, $m=r-(2d-1)$, such that ${\mathcal{K}}(H_d)+\operatorname{span}\{W_{a_\ell b_\ell}\}$ has dimension $r$. This is possible because the $W_{ab}$ span ${M_d(\mathbb{C})}$. Then $\tfrac12H_d+\tfrac{1}{2m}\sum_\ell W_{a_\ell b_\ell}^*(\cdot)W_{a_\ell b_\ell}$ is a Markov operator with Kraus space of dimension $r$ containing ${\mathcal{K}}(H_d)$, hence positivity improving by Theorem~\ref{thm:kraus}.
\end{proof}

\begin{proposition}[Hankel channel]\label{prop:hankel}
For $k=2,\dots,2d$ let $J_k:=\sum_{i+j=k}E_{ij}$ (anti-diagonal partial permutations, indices $1\le i,j\le d$), and define
\[
H_d(a):=\frac1d\sum_{k=2}^{2d}J_k\,a\,J_k,\qquad a\in{M_d(\mathbb{C})} .
\]
Then $H_d$ is a bistochastic Markov operator ($H_d$ and $(H_d)_*=H_d$ are both unital). Its Kraus space is the space of Hankel matrices, its Kraus rank is $2d-1$, and it is positivity improving. Hence the bound of Corollary~\ref{cor:rank}(i) is attained in every dimension.
\end{proposition}

\begin{proof}
Each $J_k$ is real symmetric, so $J_k^*=J_k$. Moreover $J_k^*J_k=\sum_{j:\,1\le k-j\le d}E_{jj}$. For fixed $j$, the indices $k$ with $1\le k-j\le d$ are exactly $k=j+1,\dots,j+d$, so $\frac1d\sum_kJ_k^*J_k={\mathbf{1}}$. Since $J_kJ_k^*=J_k^*J_k$, also $(H_d)_*({\mathbf{1}})={\mathbf{1}}$. The $J_k$ have disjoint supports, so they are linearly independent and span the Hankel matrices; thus $r(H_d)=2d-1$.
For positivity improvement we use Theorem~\ref{thm:kraus}(ii). Identify $\varphi=\sum\varphi_if_i$ and $\psi=\sum\psi_jf_j$ with the polynomials $\tilde\varphi(t)=\sum_i\overline{\varphi_i}t^i$ and $\tilde\psi(t)=\sum_j\psi_jt^j$. Then
\[
{\langle \varphi, J_k\psi\rangle}=\sum_{i+j=k}\overline{\varphi_i}\psi_j
\]
is the coefficient of $t^k$ in $\tilde\varphi\tilde\psi$. If all these coefficients vanished, then $\tilde\varphi\tilde\psi=0$ in ${\mathbb{C}}[t]$, and since ${\mathbb{C}}[t]$ is an integral domain, $\varphi=0$ or $\psi=0$.
\end{proof}

In the language of \cite{Par04,Bha06}, $V({\mathcal{K}}(H_d)^\perp)$ is a completely entangled subspace of the maximal dimension $(d-1)^2$. It consists of the matrices all of whose anti-diagonal sums vanish.

\section{Classification in Dimension Two}\label{sec:d2}

\begin{theorem}[Qubits]\label{thm:d2}
Let $P$ be a Markov operator on $M_2({\mathbb{C}})$ with Kraus rank $r=r(P)$.
\begin{enumerate}[label=(\roman*)]
\item If $r\le2$, then $P$ is not positivity improving.
\item If $r=4$, then $P$ is positivity improving.
\item If $r=3$, let ${\mathcal{K}}(P)^\perp={\mathbb{C}} Z$ with $Z\ne0$. Then $P$ is positivity improving if and only if $\det Z\ne0$, equivalently, if and only if $\ker C_P$ is spanned by an entangled vector (of Schmidt rank two).
\end{enumerate}
\end{theorem}

\begin{proof}
A nonzero $2\times2$ matrix has rank one if and only if it is singular. Hence, by Theorem~\ref{thm:kraus}, $P$ is positivity improving if and only if every nonzero $Z\in{\mathcal{K}}(P)^\perp$ is invertible.
(i) If $r\le2$, then $\dim{\mathcal{K}}(P)^\perp\ge2$. Choose linearly independent $Z_1,Z_2\in{\mathcal{K}}(P)^\perp$. The function $q(s,t)=\det(sZ_1+tZ_2)$ is a binary quadratic form (possibly degenerate or identically zero), so it has a zero $(s,t)\ne(0,0)$ over ${\mathbb{C}}$. Then $sZ_1+tZ_2$ is a nonzero singular element of ${\mathcal{K}}(P)^\perp$.
(ii) Here ${\mathcal{K}}(P)^\perp=\{0\}$.
(iii) The nonzero elements of ${\mathcal{K}}(P)^\perp$ are the multiples $cZ$, $c\ne0$, which are invertible if and only if $Z$ is. The Schmidt rank of $V(Z)$ equals ${\operatorname{rank}} Z$.
\end{proof}

\noindent Thus, for qubits, the minimal Kraus rank $3=2d-1$ is necessary, but it is sufficient only under the nondegeneracy condition $\det Z\ne0$. The following family shows that all cases occur and that the estimates of Theorem~\ref{thm:ergodic} are optimal.

\begin{proposition}[Pauli channels]\label{prop:pauli}
Let $\sigma_0={\mathbf{1}}$, $\sigma_1,\sigma_2,\sigma_3$ be the Pauli matrices and let ${P_{*}}(x)=\sum_{j=0}^3p_j\sigma_jx\sigma_j$ with $p_j\ge0$, $\sum_jp_j=1$. For $\{j,k,l\}=\{1,2,3\}$ put $s_j:=p_0+p_j-p_k-p_l$ and $s:=\max_j|s_j|$. Then:
\begin{enumerate}[label=(\roman*)]
\item $P$ is positivity improving if and only if at least three of the $p_j$ are strictly positive, if and only if $s<1$;
\item $\lambda(P)=(1-s)/2$, $\delta(P)=s$, and ${\operatorname{spec}}({P_{*}})=\{1,s_1,s_2,s_3\}$;
\item consequently, if $P$ is positivity improving, then $\delta(P)=1-d\lambda(P)$ and $\gamma(P)=d\lambda(P)$, i.e.\ equality holds in Theorem~\ref{thm:ergodic}(i) and (iv).
\end{enumerate}
\end{proposition}

\begin{proof}
The Pauli matrices are Hilbert--Schmidt orthogonal, so $r(P)=\#\{j:p_j>0\}$, and ${\mathcal{K}}(P)$ is spanned by the $\sigma_j$ with $p_j>0$. If exactly three $p_j$ are positive, then ${\mathcal{K}}(P)^\perp$ is spanned by the remaining Pauli matrix, which is invertible. Theorem~\ref{thm:d2} gives the first equivalence in (i).
Since $\sigma_k\sigma_j\sigma_k=\sigma_j$ if $k\in\{0,j\}$ and $\sigma_k\sigma_j\sigma_k=-\sigma_j$ otherwise, ${P_{*}}(\sigma_j)=s_j\sigma_j$ for $j=1,2,3$, and ${P_{*}}({\mathbf{1}})={\mathbf{1}}$. This gives the spectrum. Writing a state as $\rho=\frac12({\mathbf{1}}+n\cdot\sigma)$ with $|n|\le1$, we get ${P_{*}}(\rho)=\frac12({\mathbf{1}}+(\Lambda n)\cdot\sigma)$ with $\Lambda=\operatorname{diag}(s_1,s_2,s_3)$. Pure states correspond to $|n|=1$, and ${\lambda_{\min}}({P_{*}}(e_\psi))=\frac12(1-|\Lambda n|)$, whose minimum over the unit sphere is $\frac12(1-s)$. Next, $\frac12\|{P_{*}}(\rho)-{P_{*}}(\rho')\|_1=\frac12|\Lambda(n-n')|\le s$, with equality for $n=-n'$ a unit vector along a direction where $|s_j|=s$. Hence $\delta(P)=s$, and $\gamma(P)=1-s$ when $s<1$. Finally, $|s_j|=1$ means $p_k=p_l=0$ (if $s_j=1$) or $p_0=p_j=0$ (if $s_j=-1$). So $s<1$ if and only if no two of the $p_j$ vanish, which completes (i). Part (iii) follows from (ii) with $d=2$.
\end{proof}

\begin{example}[Depolarizing channel]
${P_{*}}(x)=(1-p)x+p{\operatorname{Tr}}(x)\frac{{\mathbf{1}}}{2}$ is completely positive if and only if $0\le p\le\frac43$. It is the Pauli channel with $p_0=1-\frac{3p}{4}$, $p_1=p_2=p_3=\frac p4$. By Proposition~\ref{prop:pauli} it is positivity improving if and only if $0<p\le\frac43$; then $\lambda(P)=\min\{p/2,\,1-p/2\}$ and $\delta(P)=|1-p|$. For $p=\frac43$ the Kraus rank is $3$ and ${\mathcal{K}}(P)^\perp={\mathbb{C}}{\mathbf{1}}$.
\end{example}

\begin{example}[Rank three without positivity improvement]\label{ex:AD}
Let $b,c\in{\mathbb{C}}\setminus\{0\}$ with $|b|^2+|c|^2=1$, and let
\[
K_1={|f_2\rangle\langle f_1|},\qquad K_2=b{|f_1\rangle\langle f_2|},\qquad K_3=c{|f_2\rangle\langle f_2|}.
\]
Then $\sum_iK_i^*K_i={\mathbf{1}}$, $r(P)=3$, and ${\mathcal{K}}(P)^\perp={\mathbb{C}} E_{11}$ with $\det E_{11}=0$. Hence $P$ is not positivity improving; indeed ${P_{*}}(e_{f_1})=e_{f_2}$.
\end{example}

\noindent For $d=2$, the Hankel channel of Proposition~\ref{prop:hankel} is the Pauli channel with $(p_0,p_1,p_2,p_3)=(\frac14,\frac12,0,\frac14)$. Thus $\lambda(H_2)=\frac14$ and $\delta(H_2)=\gamma(H_2)=\frac12$.

\section{Classification in Dimension Three}\label{sec:d3}

For $Z\in M_3({\mathbb{C}})$ we use the classical fact that ${\operatorname{rank}} Z\le1$ if and only if all $2\times2$ minors of $Z$ vanish, i.e.\ if and only if ${\operatorname{adj}}(Z)=0$, where ${\operatorname{adj}}$ denotes the adjugate matrix.

\begin{theorem}[Qutrits]\label{thm:d3}
Let $P$ be a Markov operator on $M_3({\mathbb{C}})$ with Kraus rank $r=r(P)$, and let $Z_1,\dots,Z_m$ be a basis of ${\mathcal{K}}(P)^\perp$, $m=9-r$.
\begin{enumerate}[label=(\roman*)]
\item If $r\le4$, then $P$ is not positivity improving.
\item If $r=9$, then $P$ is positivity improving.
\item If $r=8$, then $P$ is positivity improving if and only if ${\operatorname{rank}} Z_1\ge2$, i.e.\ ${\operatorname{adj}}(Z_1)\ne0$.
\item If $5\le r\le7$, then $P$ is positivity improving if and only if the nine quadratic forms
\[
q_{ab}(t_1,\dots,t_m):=\Big[{\operatorname{adj}}\Big(\sum_{\ell=1}^mt_\ell Z_\ell\Big)\Big]_{ab},\qquad 1\le a,b\le3,
\]
have no common zero in ${\mathbb{C}}^m\setminus\{0\}$. For $r=7$ ($m=2$) this holds if and only if the binary forms $q_{ab}$ are not all identically zero and have no common linear factor, i.e.\ their greatest common divisor is constant.
\end{enumerate}
\end{theorem}

\begin{proof}
By Theorem~\ref{thm:kraus}, $P$ is positivity improving if and only if every nonzero $Z\in{\mathcal{K}}(P)^\perp$ satisfies ${\operatorname{rank}} Z\ge2$, i.e.\ ${\operatorname{adj}}(Z)\ne0$. Writing $Z=\sum_\ell t_\ell Z_\ell$ with $t\ne0$ gives (iii) and the first part of (iv).
(i) is Corollary~\ref{cor:rank}(i) with $d=3$, since $2d-1=5$. (ii) is Corollary~\ref{cor:rank}(ii).
For $m=2$, each $q_{ab}$ is a binary quadratic form or zero. Over ${\mathbb{C}}$, a family of binary forms has a common zero in ${\mathbb{C}}^2\setminus\{0\}$ if and only if all of them vanish there, which happens if and only if they share a linear factor or all vanish identically.
\end{proof}

\begin{remark}\label{rem:wrong}
(a) Theorem~\ref{thm:kraus} shows that, in every dimension, the spanning condition ``$\{K_i\psi\}$ spans ${\mathcal{H}}$ for all $\psi\ne0$'' is by itself equivalent to positivity improvement. Irreducibility is then automatic by Proposition~\ref{prop:struct}(iii), so it is not an independent condition.
(b) For $d=3$ there is no positivity improving Markov operator with three Kraus operators. Indeed, $\psi\mapsto\det[K_1\psi\ K_2\psi\ K_3\psi]$ is a homogeneous cubic form on ${\mathbb{C}}^3$, and every such form has a nontrivial zero, at which $K_1\psi,K_2\psi,K_3\psi$ fail to span ${\mathbb{C}}^3$. This is a special case of Theorem~\ref{thm:d3}(i).
(c) The criterion of Theorem~\ref{thm:d3}(iv) is decidable by elimination theory (resultants or Gr\"obner bases) in a fixed number of variables, in contrast with the NP-hardness for general $d$ \cite{GQ17}.
\end{remark}

\begin{example}[Werner--Holevo channels]\label{ex:WH}
Let $d\ge2$ and let
\[
W^{\pm}_*(x)=\frac{{\operatorname{Tr}}(x){\mathbf{1}}\pm x^T}{d\pm1}
\]
be the Werner--Holevo channels \cite{WH02}. From $\sum_{j,k}E_{jk}\otimes E_{kj}=F$ (the flip) one gets $C_{W^\pm}=\frac{2}{d\pm1}\Pi_\pm$, where $\Pi_+$ and $\Pi_-$ are the projections onto the symmetric and antisymmetric subspaces. Hence $W^\pm$ are Markov operators, and $V^{-1}(\operatorname{ran}\Pi_\pm)$ identifies ${\mathcal{K}}(W^+)$ with the symmetric matrices and ${\mathcal{K}}(W^-)$ with the antisymmetric matrices.

(a) ${\mathcal{K}}(W^+)^\perp$ consists of antisymmetric matrices, which have even rank, so $W^+$ is positivity improving, with $r(W^+)=d(d+1)/2$ ($=6$ for $d=3$). Directly, $W^+_*(e_\psi)=({\mathbf{1}}+e_{\bar\psi})/(d+1)$, so $\lambda(W^+)=1/(d+1)$. Since $W^+_*(\rho-\sigma)=(\rho-\sigma)^T/(d+1)$ and the transpose is a trace-norm isometry, $\delta(W^+)=1/(d+1)=1-d\lambda(W^+)$. The restriction of $W^+_*$ to traceless matrices is $x\mapsto x^T/(d+1)$, with eigenvalues $\pm1/(d+1)$, so $\gamma(W^+)=d/(d+1)=d\lambda(W^+)$. All bounds of Theorem~\ref{thm:ergodic} are therefore attained in every dimension.

(b) $W^-$ is never positivity improving, since $W^-_*(e_\psi)\bar\psi=0$. For $d=3$, $r(W^-)=3<5$.
\end{example}

\begin{example}[Weyl channels on ${\mathbb{C}}^3$]\label{ex:weyl}
Let $\omega=e^{2\pi i/3}$, let $X f_j=f_{j+1}$ and $Zf_j=\omega^{j}f_j$ (indices mod $3$), and let $W_{ab}=X^aZ^b$. For a probability vector $(p_{ab})$ put ${P_{*}}(x)=\sum_{a,b}p_{ab}W_{ab}xW_{ab}^*$. Since the $W_{ab}$ are orthogonal, ${\mathcal{K}}(P)=\operatorname{span}\{W_{ab}:p_{ab}>0\}$ and $r(P)=\#\{(a,b):p_{ab}>0\}$.
\begin{itemize}
\item $r=9$: positivity improving (completely depolarizing channel for $p_{ab}\equiv\frac19$).
\item $r=8$, $p_{ab}=0$ for exactly one $(a,b)$: ${\mathcal{K}}(P)^\perp={\mathbb{C}} W_{ab}$ with $W_{ab}$ unitary, so $P$ is positivity improving. For uniform weights and $(a,b)=(0,0)$, ${P_{*}}(x)=\frac18(3{\operatorname{Tr}}(x){\mathbf{1}}-x)$ and $\lambda(P)=\frac14$.
\item $r=7$, $p_{01}=p_{02}=0$: ${\mathcal{K}}(P)^\perp=\operatorname{span}\{Z,Z^2\}$, and $sZ+tZ^2=\operatorname{diag}(s+t,\ s\omega+t\omega^2,\ s\omega^2+t\omega)$. Any two of the three diagonal entries vanish simultaneously only for $s=t=0$, since the corresponding $2\times2$ coefficient matrices have determinants $\omega^2-\omega$, $\omega-\omega^2$ and $\omega^2-\omega^4\neq0$. Hence every nonzero element has rank $\ge2$, and $P$ is positivity improving.
\item $r=6$, $p_{ab}>0$ if and only if $a\ne0$: ${\mathcal{K}}(P)^\perp=\operatorname{span}\{{\mathbf{1}},Z,Z^2\}$ is the diagonal algebra, which contains $E_{11}$. So $P$ is \emph{not} positivity improving; indeed ${P_{*}}(e_{f_1})$ is supported on $\operatorname{span}\{f_2,f_3\}$. Together with Example~\ref{ex:WH}(a), this shows that for $d=3$ and $r=6$ both alternatives occur.
\item $r=3$: the dephasing channel ($p_{0b}=\frac13$) and the cyclic twirl ($p_{a0}=\frac13$) are not positivity improving. For the cyclic twirl ${P_{*}}(x)=\frac13\sum_aX^axX^{-a}$, the Fourier vectors $\psi$ (eigenvectors of $X$) satisfy ${P_{*}}(e_\psi)=e_\psi$.
\end{itemize}
\end{example}

\begin{example}[Minimal rank for qutrits]
The Hankel channel $H_3$ of Proposition~\ref{prop:hankel} has Kraus rank $5$ and is positivity improving. Its Kraus operators are $\frac{1}{\sqrt3}J_k$, $k=2,\dots,6$, where $J_2=E_{11}$, $J_3=E_{12}+E_{21}$, $J_4=E_{13}+E_{22}+E_{31}$, $J_5=E_{23}+E_{32}$ and $J_6=E_{33}$. The complement ${\mathcal{K}}(H_3)^\perp$ is the four-dimensional space of matrices with vanishing anti-diagonal sums. For instance, $Z=E_{12}-E_{21}$ and $Z=E_{13}-E_{22}$ have rank $2$, in agreement with Theorem~\ref{thm:d3}(iv). Numerically one finds $\lambda(H_3)\approx0.0370$, so Theorem~\ref{thm:ergodic} yields $\delta(H_3)\le 1-3\lambda(H_3)\approx0.889$.
\end{example}

\section{Remarks on the Infinite-Dimensional Case}\label{sec:infinite}

Let now ${\mathcal{H}}$ be an infinite-dimensional separable Hilbert space, let $P$ be a normal unital completely positive map on ${\mathcal{B}(\mathcal{H})}$, and let ${P_{*}}$ be its predual on the trace class ${\mathcal{T}(\mathcal{H})}$. Since positive trace-class operators cannot be boundedly invertible, the appropriate replacement for strict positivity is \emph{faithfulness} (trivial kernel). We say that $P$ is positivity improving if $P(a)$ is faithful for every nonzero $a\ge0$ in ${\mathcal{B}(\mathcal{H})}$. By trace duality and testing rank-one projections, this is equivalent to faithfulness of ${P_{*}}(x)$ for every nonzero positive trace-class $x$.

\begin{proposition}\label{prop:inf}
\begin{enumerate}[label=(\roman*)]
\item If ${P_{*}}(x)=\sum_iK_ixK_i^*$ with a trace-norm convergent Kraus series for trace-class $x$, then $P$ is positivity improving if and only if for all nonzero $\varphi,\psi\in{\mathcal{H}}$ there is $i$ with ${\langle \varphi, K_i\psi\rangle}\ne0$.
\item Theorem~\ref{thm:MDineq} holds verbatim: if ${P_{*}}(\rho)\ge\kappa\sigma$ for all normal states $\rho$, then $\delta(P)\le1-\kappa$.
\item The infimum of the spectrum of every positive trace-class output in infinite dimension is zero; hence the direct spectral analogue of $\lambda(P)$ vanishes. Moreover, positivity improvement does not imply $\delta(P)<1$.
\end{enumerate}
\end{proposition}

\begin{proof}
(i) For $x\ge0$ nonzero in ${\mathcal{T}(\mathcal{H})}$, the operator ${P_{*}}(x)$ dominates a positive multiple of ${P_{*}}(e_\psi)$ for an eigenvector $\psi$ of $x$ with eigenvalue $\|x\|$. As in Proposition~\ref{prop:dual}, faithfulness of all ${P_{*}}(x)$ is equivalent to ${\langle \varphi, {P_{*}}(e_\psi)\varphi\rangle}=\sum_i|{\langle \varphi, K_i\psi\rangle}|^2>0$ for all nonzero $\varphi,\psi$.
(ii) The proof of Theorem~\ref{thm:MDineq} uses only positivity and trace preservation.
(iii) ${P_{*}}(e_\psi)$ is a compact positive operator, so ${\lambda_{\min}}$ must be replaced by $\inf{\operatorname{spec}}$, which is $0$. For the second assertion, let $(f_n)_{n\ge1}$ be an orthonormal basis, let $\tau$ be a faithful normal state, let $0<\varepsilon_n\to0$, and put $\sigma_n=(1-\varepsilon_n)e_{f_n}+\varepsilon_n\tau$. The measure-and-prepare map
\[
{P_{*}}(x)=\sum_{n\ge1}{\langle f_n, xf_n\rangle}\,\sigma_n
\]
is a normal quantum channel. For $\psi\ne0$ we have ${P_{*}}(e_\psi)\ge\big(\sum_n|{\langle f_n, \psi\rangle}|^2\varepsilon_n\big)\tau$, which is faithful because the coefficient is strictly positive. Hence $P$ is positivity improving. On the other hand,
\[
\tfrac12\|{P_{*}}(e_{f_n})-{P_{*}}(e_{f_m})\|_1=\tfrac12\|\sigma_n-\sigma_m\|_1\ge1-\varepsilon_n-\varepsilon_m\longrightarrow1\quad(n\neq m\to\infty),
\]
so $\delta(P)=1$.
\end{proof}

Thus, in infinite dimensions, the equivalence of Theorem~\ref{thm:MD} breaks down, and the Markov--Dobrushin condition with a faithful reference state must be imposed as a genuinely stronger hypothesis. This is in line with the classical theory, where positivity improvement must be supplemented by compactness or Doeblin-type conditions to obtain uniform ergodicity \cite{Sen06,Muk13}. Infinite-dimensional transition operators of quantum Markov chains on trees and lattices \cite{ASS20,MS19}, and transfer operators of finitely correlated states \cite{FNW92}, are natural settings in which these refined conditions can be checked. This will be pursued elsewhere.

\section{Conclusion}

We have shown that positivity improvement of a quantum Markov operator on a matrix algebra is a self-dual property that coincides with the quantum Markov--Dobrushin minorization with respect to the trace. It implies an explicit and optimal contraction of the Dobrushin ergodicity coefficient, a spectral gap $\gamma(P)\ge d\lambda(P)$, and uniform exponential convergence to a unique faithful invariant state. Structurally, positivity improvement is exactly the absence of rank-one operators in the orthogonal complement of the Kraus space, i.e.\ complete entanglement of the kernel of the Choi matrix. This yields the sharp bound $r(P)\ge2d-1$, attained by the Hankel channel. For $d=2$ and $d=3$ we obtained complete and effectively checkable classifications, and the Pauli, Weyl and Werner--Holevo families show that all cases occur and that our estimates cannot be improved.



\end{document}